# The Applicability of ISO/IEC 25023 Measures to the Integration of Agents and Automation Systems

Stamatis Karnouskos[1], Roopak Sinha[2], Paulo Leitão[3], Luis Ribeiro[4], Thomas. I. Strasser[5]

[1]SAP, Walldorf, Germany, email: stamatis.karnouskos@sap.com
[2]IT & Software Engineering, Auckland University of Technology, New Zealand, email: roopak.sinha@aut.ac.nz
[3]Research Centre in Digitalization and Intelligent Robotics (CeDRI), Instituto Politécnico de Bragança, Campus de Santa Apolónia, 5300-253 Bragança, Portugal, email: pleitao@ipb.pt
[4]Linköping University, SE-58 183 Linköping, Sweden, email: luis.ribeiro@liu.se
[5]Center for Energy – AIT Austrian Institute of Technology and Institute of Mechanics and Mechatronics – Vienna University of Technology, Vienna, Austria, email: thomas.i.strasser@ieee.org

**Abstract**

*The integration of industrial automation systems and software agents has been practiced for many years. However, such an integration is usually done by experts and there is no consistent way to assess these practices and to optimally select one for a specific system. Standards such as the ISO/IEC 25023 propose measures that could be used to obtain a quantification on the characteristics of such integration. In this work, the suitability of these characteristics and their proposed calculation for assessing the connection of industrial automation systems with software agents is discussed. Results show that although most of the measures are relevant for the integration of agents and industrial automation systems, some are not relevant in this context. Additionally, it was noticed that some measures, especially those of a more technical nature, were either very difficult to computed in the automation system integration, or did not provide sufficient guidance to identify a practice to be used.*

## 1. INTRODUCTION

Industrial Agents (IA) [1] have often been used for a variety of activities in industry, including their integration with automation systems. With the emergence of Cyber-Physical Systems (CPS) and the Internet of Things (IoT) approaches such agent-based integrations are expected to be further widespread into different domains like factory automation, power & energy systems, and building automation [2]–[5].

In cybernetics [4], the integration of software and hardware systems still remains a highly challenging issue [6], [7]. There are several factors causing integration failures [8], some of which may be better tackled or even prevented by focusing on integration practices that are more suitable for the specific use case. Hence, a key issue is the lack of an easy and consistent way to assess such integration practices [9] and to select the most appropriate ones for a particular use case.

In software systems research, several quality attributes and metrics are studied [10], [11]. Standards, such as ISO/IEC 25010 [12], define several high-level characteristics which are relevant to this context [13]. However, to be able to get a quantifiable assessment, there is a need to check the concrete measures and their calculation as proposed by ISO/IEC 25023 [14] which also belongs to the same standards family. The primary contributions of this paper, therefore, is a critical discussion of the usage of those measures for assessing the integration of software agents and low-level automation functions of industrial systems.

In a CPS context, characterized by a network of entities that integrate computational and physical counterparts, IA extend the traditional characteristics of software agents, particularly intelligence, autonomy, and cooperation, with industrial requirements, namely hardware integration, reliability, faulttolerance, scalability, standard compliance, quality assurance, resilience, manageability, and maintainability [15]. In this perspective, the interface between software agents, often referred to as High-Level Control (HLC), and Low-Level Controllers (LLC) performing industrial automation functions (like Programmable Logic Controllers (PLC), Industry PCs (IPC), or robots) assumes critical importance to achieve the industrial requirements, to fully comply with the enterprise operational context and to guarantee the business continuity.

As illustrated in Figure 1, this interface usually comprises of an HLC Application Programming Interface (API), a communication channel and a LLC API [16], [17]. The agents execute in an agent platform and interact with industrial devices via those APIs. The nature of the interaction varies such as local or remote calls, publish/subscribe or client/server messages, etc. The communication channel may also differ between practices. Besides, there are several variations concerning the physical location of the agent (e.g., on-device and remotely). All practices can be to a degree mapped to this high-level integration model [9], which can be used to apply the measurable characteristics investigated in this work.

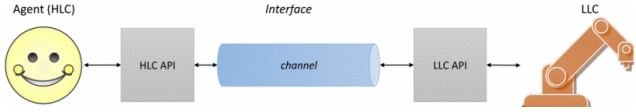

Figure 1. Industrial Agent Integration Model.

The remaining parts of this paper are organized as follows: section 2 presents the assessment of the different characteristics and sub-characteristics established by ISO/IEC 25023 to be applied in the software interface context. section 3 discusses the main findings of this work and finally, section 4 rounds up the paper with conclusions and future works.

## 2. ASSESSMENT OF ISO/IEC 25023 SUITABILITY

Table I presents an overview of the different characteristics, sub-characteristics, measures and finally the assessment of each measure (discussed below) with respect to its suitability in the integration of software a and low-level automation functions. To assess the level of suitability of each measure, a Likert scale with the following coding is being used: Very Good, Good, Neutral, Poor, and Very Poor.

In the following the different characteristics and sub-characteristics are being analyzed regarding their suitability to assess the integration of software agents with low-level automation functions.

### A. Functional Suitability

Functional suitability refers to the degree *"to which a product or system provides functions that meet stated and implied needs when used under specified conditions"* [12]. This characteristic is critical to ensure the proper operation of the interface integrating IA and low-level automation functions.

*1) Functional Completeness:* This sub-characteristic can be measured by considering the system *functional coverage*, that is the metric of how much the specified and designed functionalities have been implemented and covered by the testbench. This measure applies to the assessment of interfacing practices, contributing to determine the percentage of specified functions that the interface implements.

*2) Functional Correctness:* It refers to the input-output behavior of an algorithm or system, i.e., the expected output produced by the system for a specific input. This measure is useful to assess the IA interfacing practices since it can indicate the percentage of functions and services implemented by the interface that provides the correct results, i.e., are working correctly.

*3) Functional Appropriateness:* Two different measures can be considered in the assessment of the functional appropriateness of an interface practice. Firstly, the *functional appropriateness of the usage objective* refers to what portion of the functions required by the user provides the appropriate outcome to achieve a specific usage objective, and secondly, the *functional appropriateness of the system* refers to what proportion of the functions required by the users to achieve their objectives provides an appropriate outcome. Both measures apply to assess the interfacing practices.

### B. Performance Efficiency

Performance efficiency enables assessing *the performance relative to the amount of resources used under stated conditions* and may include software products, system configuration or materials [12]. Performance metrics are of paramount importance when assessing an integration practice in industrial automation, as failing to meet the performance requirements voids the validity of the practice itself.

*1) Time behavior:* Each practice should be able to meet certain timing objectives like *mean response time*, in terms of first response, or *mean turnaround time*, in respect to full task completion. These measurements are generally highly suitable. The adequacy of a practice with respect to *response time adequacy* or *turnaround time adequacy* needs to be assessed in two distinct scenarios. Hard-real-time capable practices in industrial automation should offer 100% adequacy. However, in soft-real-time applications, deviations are inherent and adequacy measurements are a proper indicator of the practice's capabilities. In these cases, the rate of task completion (*mean throughput*) is also variable and the number of tasks completed for a reference time frame may provide a complementary adequate performance measurement. Nevertheless, the requirements and suitability of the different time behavior related metrics are inherently bound to the application domain where the practice is going to be operationalized.

*2) Resource Utilization:* Resource utilization metrics such as *mean processor, memory, I/O device utilization, and bandwidth utilization* indicate the consumption of CPU time, memory, and bandwidth for given performance measures. Bandwidth utilization can, in principle, be easily quantified for each practice. Quantifying CPU time and memory footprint is challenging as these require isolating agent processes and low-level aspects of a practice; this is very difficult for multi-core systems using modern self-optimizing operating systems. While on the agent side it may be possible to isolate these processes, on the side of the low-level controller the variety of implementation options increases the variability of the results.

*3) Capacity:* Capacity measurements such as *transaction processing capacity*, *user access capacity* and *user access increase adequacy* are related to the ability of the system to accommodate simultaneous user access. Typically, a practice, as discussed in this context, has one specific user, the agent, part of which is included in the practice itself. In case of increasing the number of agents (e.g., in a multi-agent system) that are operated by different users, and compete for the same goal (e.g., access to the automation device), the case is more complex, and the proposed *user access increase adequacy proposed* measure becomes relevant.

**C. Compatibility**

Compatibility measurements *"assess the degree to which a product, system or component can exchange information with other products, systems or components, and perform its required functions while sharing the same hardware or software environment"* [12].

*1) Co-existence:* Co-existence tries to capture the ability to mix software on the same computational platform. Well-designed software should be able to co-exist without interfering with the operation of other running systems on the same computational platform. Certain classes of software are however noticeably known for not tolerating the presence of similar software. Typical examples are anti-virus, security suites, backup or software upgrade processes. In the discussed context, properly designed practices must be able to coexist given sufficient computational resources. However, it is generally very challenging to characterize co-existence in a proper way in the discussed agent integration case.

*2) Interoperability:* Interoperability is fundamental in software integration efforts, especially when it pertains to complex industrial devices. Interoperable systems successfully connect among them following a well defined and compatible set of interaction protocols, syntax and data formats, that are minimal and essential to enable them to interact. The measures defined in [12] are seen as partly appropriate, at least from their description, as they refer to *data format exchangeability* and *data exchange protocol sufficiency*. For agent and industrial system integrations, more than one data format may exist, but it is not a common case. Similarly for data exchange protocols, usually only one protocol is available, over which the data is exchanged during the interaction. Therefore, the proposed measurement function, which is a ratio of supported over the total number of formats or protocols, is not seen as meaningful for very focused small systems such as the ones which are addressed here. Adding more for example protocols or data formats to a practice may increase the measures as currently defined in [12], but does not provide adequate info on the quality of these. In addition, doing so may be considered as establishing a new practice whose characteristics need to be analyzed for that specific instantiation. In that case interoperability measurements as defined in [12] hardly apply (since most practices feature a single data exchange protocol and a single data format). In addition, the *external interface adequacy* is not seen as appropriate, as usually external interfaces are exposed only when they are to be used, but not otherwise, as typical interaction processes in automation systems are well-defined.

**D. Usability**

Usability measures ensure that product usage can be easily understood, learned and operated. As usability is highly subjective, a representative and sufficiently large group of user feedback needs to be obtained. The suitability of the ISO/IEC 25023 proposed measures varies, and although description-wise some would be relevant, the way ISO/IEC

Table I. APPLICABILITY OF ISO/IEC 2503 MEASURES IN THE IA CONTEXT.

| Characteristics | Sub-characteristics | Measure | IA suitability |
|---|---|---|---|
| Functional Suitability | Functional Completeness | Functional coverage | Good |
| | Functional Correctness | Functional correctness | Very Good |
| | Functional Appropriateness | Functional appropriateness of usage objective | Very Good |
| | | Functional appropriateness of the system | Good |
| Performance Efficiency | Time behavior | Mean response time | Very Good |
| | | Response time adequacy | Good |
| | | Mean turnaround time | Very Good |
| | | Turnaround time adequacy | Good |
| | | Mean throughput | Very Good |
| | Resource Utilization | Mean processor utilization | Neutral |
| | | Mean memory utilization | Neutral |
| | | Mean I/O devices utilization | Neutral |
| | | Bandwidth utilization | Good |
| | Capacity | Transaction processing capacity | Poor |
| | | User access capacity | Poor |
| | | User access increase adequacy | Poor |
| Compatibility | Co-existence | Co-existence with other products | Poor |
| | Interoperability | Data formats exchangeability | Neutral |
| | | Data exchange protocol sufficiency | Neutral |
| | | External interface adequacy | Poor |
| Usability | Appropriateness recognisability | Description completeness | Good |
| | | Demonstration coverage | Neutral |
| | | Entry point self-descriptiveness | Very Poor |
| | Learnability | User guidance completeness | Very Good |
| | | Entry fields defaults | Very Good |
| | | Error messages understandability | Very Good |
| | | Self-explanatory user interface | Good |
| | Operability | Operational consistency | Very Good |
| | | Message clarity | Very Good |
| | | Functional customizability | Good |
| | | User interface customizability | Poor |
| | | Monitoring capability | Very Good |
| | | Undo capability | Very Good |
| | | Understandable categorization of information | Very Good |
| | | Appearance consistency | Good |
| | | Input device support | Good |
| | User error protection | Avoidance of user operation error | Very Good |
| | | User entry error correction | Very Good |
| | | User error recoverability | Very Good |
| | User interface aesthetics | Appearance aesthetics of user interfaces | Very Poor |
| | Accessibility | Accessibility for users with disabilities | Very Poor |
| | | Supported languages adequacy | Very Poor |
| Reliability | Maturity | Fault correction | Poor |
| | | Mean time between failure (MTBF) | Very Good |
| | | Failure rate | Very Good |
| | | Test coverage | Poor |
| | Availability | System availability | Very Good |
| | | Mean down time | Very Good |
| | Fault tolerance | Failure avoidance | Good |
| | | Redundancy of components | Neutral |
| | | Mean fault notification time | Very Good |
| | Recoverability | Mean recovery time | Very Good |
| | | Backup data completeness | Poor |
| Security | Confidentiality | Access controllability | Very Good |
| | | Data encryption correctness | Neutral |
| | | Strength of cryptographic algorithms | Neutral |
| | Integrity | Data integrity | Very Good |
| | | Internal data corruption prevention | Good |
| | | Buffer overflow prevention | Neutral |
| | Non-repudiation | Digital signature usage | Good |
| | Accountability | User audit trail completeness | Neutral |
| | | System log retention | Very Good |
| | Authenticity | Authentication mechanism sufficiency | Very Good |
| | | Authentication rules conformity | Very Good |
| Maintainability | Modularity | Coupling of components | Very Good |
| | | Cyclomatic complexity adequacy | Neutral |
| | Reusability | Reusability assets | Very Good |
| | | Coding rules conformity | Good |
| | Analyzability | System log completeness | Very Good |
| | | Diagnosis function effectiveness | Poor |
| | | Diagnosis function sufficiency | Poor |
| | Modifiability | Modification efficiency | Neutral |
| | | Modification correctness | Very Good |
| | | Modification capability | Neutral |
| | Testability | Test function completeness | Good |
| | | Autonomous testability | Very Good |
| | | Test restartability | Neutral |
| Portability | Adaptability | Hardware environmental adaptability | Very Good |
| | | System software environmental adaptability | Good |
| | | Operational environment adaptability | Very Good |
| | Installability | Installation time efficiency | Good |
| | | Ease of installation | Very Good |
| | Replaceability | Usage similarity | Very Good |
| | | Product quality equivalence | Good |
| | | Functional inclusiveness | Good |
| | | Data reusability/import capability | Neutral |

25023 proposes to measure them may not be suitable for the context we investigate in this work.

*1) Appropriateness recognizability:* Users need to be able to easily recognize and match product descriptions,

demonstration features and other capabilities to their purposes. To do so, *description completeness*, *demonstration coverage* and *entry point self-descriptiveness* measures are defined as the ratios of what is explained to what the product is capable of. While these measures help recognize the product capabilities, for industrial integration such description could be done directly using code examples and manuals. Entry point self-descriptiveness is linked to the product website and may not be very relevant as a low-level integration measure.

*2) Learnability:* The ease at which we can learn to use a product is linked to appropriateness recognizability. The standard defines several measures to assess *learnability*. *User guidance completeness* is highly relevant as for software and hardware integration as is our case. Similarly, *entry fields defaults* empower the users and simplify integration by utilizing meaningful default values that subsequently the user can change. Understanding why an error has happened is measured in *error messages understandability* and is important for complex systems that span both software and hardware. *Self-explanatory user interface* addresses the need for an easy understanding of the elements presented to the user, and while most current implementations do not usually feature a user interface, this is seen also as a must to ease user on-boarding.

*3) Operability:* Several measures are as the ease with which the product is operated/controlled. The consistent behavior measured by *operational consistency* is a must for industrial systems that need a reliable and deterministic execution pattern. Equally important is *message clarity* that assesses the correctness of instructions from the product to the user, as in industry ambiguous instructions may have far-reaching effects, especially in critical systems domain. In this context also *understandable categorization of information* measure is suitable, in order to enable easy and correct information categorization, e.g., if an alarm is raised and what its severity level might be. Customization of functions and user interface as measured by *functional customizability* is seen as highly relevant. Aspects relevant to the user interface such as *user interface customizability* and *appearance consistency* are suitable measures but of lesser importance mainly to the currently wide-spread lack of user interfaces for agent and low-level automation function practices. Measuring *monitoring capability* is seen as extremely important, as any deployment in industrial environments needs to be subjected to a wide range of third-party infrastructure and process monitoring tools. The *undo capability* has also its merits, in order to be able to go back to a previous functional state that reverses the changes introduced in the system and which may have unwanted effects, usually discovered only during operation stage. The *input device support* measures the extent to which tasks can be initiated by other input modalities, and is very suitable as the agent solution needs to interact with other systems.

*4) User Error Protection*: Protecting the users from making errors during the operational stage is a highly relevant quality as they protect both the system and the users. The *avoidance of user operation error* measure is suitable, but its assessment may imply reverting to interactive mode, e.g., to ask for operator confirmation. Being able to recognize potential malfunctioning due to a false user entry, and propose a correct value with appropriate justification is assessed in *user entry error correction* is very suitable for agent-based integrations. Similarly important is to be able to recover from a user error, and as so *user error recoverability* is seen as key. In industrial settings, proper testing minimizes the possibility for such errors, and the interaction with the user is limited. However, when things go wrong, and no such error protection measures are in place, cascaded effects may be horrendous.

*5) User Interface Aesthetics:* Aesthetic satisfaction of the user is addressed in *appearance aesthetics of user interfaces*. However, as already mentioned due to the general lack of user interfaces when it comes to the agent and industrial system integration, this measure is of low importance.

*6) Accessibility:* Similar to the aesthetics, accessibility as measured in *accessibility for users with disabilities* and *supported languages adequacy* are not seen as relevant due to the general lack of user interaction, as agent and industrial system integration relies mostly on automated interactions, and not really on human users.

**E. Reliability**
Reliability refers to the degree *"to which a system, product or component performs specified functions under specified conditions for a specified period of time"* [12]. In industrial environments, the reliability of the interface between IA and low-level automation is so much more important as greater is the criticality of the application.

*1) Maturity:* The *Mean Time Between Failure* (MTBF) is the predicted elapsed time between inherent failures of an interface practice during its normal operation, and can be calculated as the average time between failures. A higher MTBF indicates that an interface practice works longer before failing, increasing its maturity and consequently its reliability. The *failure rate* provides the frequency with which the interface practice fails, expressed in failures per unit of time. The failure rate and MTBF measures play an important role to assess the different interface practices. Two other measures are proposed by ISO/IEC 25023, but are more related to the design, coding and testing phases. The *fault correction* measure indicates the proportion of detected faults that have been corrected during the design/coding/testing phase, and the *test coverage* measure indicates the percentage of the interface practice capabilities or functions that are executed when a particular test suite runs. An interface practice with higher test coverage has more of its functions executed during the testing phase, which suggests a lower possibility of presenting undetected bugs when compared with a practice with a lower test coverage. However, when aiming to assess the operation phase of an interface practice, these last two measures are of low relevance.

*2) Availability:* The *system availability* measure provides an indication of the percentage of the time that the interface practice is actually available over the scheduled operational time, translated by the probability that the interface practice is functioning when needed, under normal operating conditions. The *Mean Down Time* (MDT) measure is the average time that the interface practice stays non-operational (unavailable). The downtime appears after the occurrence of a failure and includes all downtime associated with repair, corrective and preventive maintenance, and logistic delays to correct the system in order to be operational again. A lower MDT means better availability and consequently better reliability of the interface practice.

*3) Fault Tolerance:* An interface practice can be tested by injecting some faults or undesired signals and see how the interface practice reacts to them. Additionally, the *failure avoidance* measure indicates the degree of mitigation of fault patterns, i.e., the percentage of fault patterns brought under control to avoid critical and serious failures. In critical systems, aiming to increase the reliability of the system, some critical parts should be redundant. The *redundancy of components* measure shows the percentage of system components that need to be installed redundantly to avoid the system failure. The fault tolerance is also improved, depending on the fastness of the notification of faults. In particular, the *Mean Fault Notification Time* (MFNT) measure indicates how fast the system reports the occurrence of faults (the closer to zero, the better the system fault tolerance is). These measures are adequate to assess the interfacing practices, but the redundancy of components measure is not critical in the interface practice context.

*4) Recoverability:* The *Mean Recovery Time* (MRT) measure is the average time that the interface practice will take to recover from failure(s). The lower the MRT is, the better the reliability of the interface practice is. In case of redundant systems, the MRT is zero (or close to it) since the redundant components (that won't exhibit failure) can take over the instant the primary one fails. However, one has to consider that for this, a switchover overhead from the primary one to the redundant one might be evident. This measure, therefore, applies very well to the assessment of the interfacing practices. Another measure of the recoverability of a system is to analyze how data is periodically backed up and can be restored in case of failure. The *Backup Data Completeness* (BDC) measure indicates the percentage of data items that are backed up regularly. In the interface practice context, this measure is not seen as suitable.

**F. Security**

Security relates to the protection of data, networks, and hardware, with data security being most important. In this context, systems containing industrial agents integrated with low-level control functions require that the interaction between them is secure. Moreover, they may also communicate securely with other systems (supervisory control, visualization, etc.).

*1) Confidentiality:* Confidentiality requires data accesses to be restricted to only authorized users or components, and to do so, the standard proposes three confidentiality measures. *Access controllability* is computed as the proportion of data items requiring access control that can only be accessed with proper authorization. *Data encryption correctness* relates to the proportion of data items that are correctly encrypted or decrypted within the system. Finally, *strength of cryptographic algorithm* relates to the proportion of protected items for which adequately strong cryptographic algorithms have been chosen. *Access controllability* is arguably the most relevant measure since IA and low-level control functions must adhere to the explicit or implicit scope of information access. Data encryption correctness and the type of cryptographic algorithms utilized are not so useful, often because these measures are met using means like secure networks, or through allowing access between components on one side (e.g., the shop floor) and another (e.g., an Internet service) through well-defined and secure interfaces. During the software design of such systems, confidentiality concerns are often deferred to the deployment stage when one or more of these means for ensuring confidentiality can be selected.

*2) Integrity:* Integrity is concerned with the prevention of unauthorized access and subsequent modification of data or program code. The standard describes three measures to assess integrity. *Data integrity* is computed as the proportion of data items to be protected that remain uncorrupted. *Internal data corruption prevention* is the proportion of available and recommended prevention methods that have been implemented into the system. Finally, *buffer overflow prevention* is the proportion of memory accesses driven through user input that are bounds checked to ensure data integrity. In this work's context, integrity assessment is very important but is carried out quite late during the development of systems containing software agents and low-level control. Buffer overflow prevention, and to some extent, data integrity, can be tested before the system is deployed. Internal data corruption prevention can be checked just before deployment through a checklist of prevention mechanisms derived from a target security standard.

*3) Non-Repudiation:* Non-repudiation is the degree to which a coherent, consistent and immutable log of events can be maintained during system execution, often for diagnostics purposes. The standard presents a single measure, i.e., *digital signature usage*, which is the proportion of events that are captured using a digital signature, certificates or other security algorithms. This is an important measure in isolating faults and identifying causality within interfaces. This measure can be strengthened through the use of blockchains, which provide a decentralized and immutable record of events, and can act as an alternative to digital signatures.

*4) Accountability:* Similar to non-repudiation, accountability also relates to tracing back an action in the system to the entity that took that action. Likewise, this is important in dynamic systems containing software agents and low-level control. The standard proposes two measures: *user audit trail completeness* and *system log retention*, each relating to keeping a log of user accesses and system actions for a suitable retention period, respectively. In the paper's context, user access logs can be handled by external interfaces, such as the standardized human-machine interface provided by systems like supervisory control. Keeping system logs is more important, and tricky, as the system can contain a multitude of software agents and low-level control functions.

*5) Authenticity:* Authenticity relates to ensuring that a resource within the system can be proved to be authentic. The standard measure authenticity through *authentication mechanism sufficiency* and *authentication rules conformity*. The former is the proportion of authentication mechanisms required in the system which have been implemented, and the latter is measured as the proportion of specified authentication rules that have been implemented. In our context, both measures can be encoded within the interfaces between software agents and low-level control. Moreover, both are equally important, as the system is dynamic by design and communication between these entities may change as time progresses.

**G. Maintainability**

Maintainability refers to the ease at which a system can be modified after it has been operationalized. It includes the on-going assessment and replacement of individual components, as well as the addition of new components to the system.

*1) Modularity:* More modular systems are easier to maintain as the addition of a component or an update to an existing one tends to have a limited impact on the rest of the system. ISO/IEC 25023 measures modularity via *coupling of components*, which is the ratio of completely independent components in a system to the number of components that must be independent. An independent component has zero impact on other parts of the system during maintenance. The standard also adds another software-specific measure, relating to the *adequacy of cyclomatic complexity*, computed as the proportion of all software modules that have an acceptable cyclomatic complexity. Acceptable cyclomatic complexity depends on the type of system. For this paper's context, coupling is an extremely important metric to capture maintainability via modularity. The highly dynamic nature of agent-control integrations requires low (or ideally zero) coupling. Cyclomatic complexity adequacy is not so important here, as often software modules within interfaces are built through the reuse of smaller, well-tested software functions.

*2) Reusability:* The ability to use one asset (hardware or software) in another system or sub-system is at the heart of most industrial systems. ISO/IEC 25023 measures it through two units: *reusability of assets* and *coding rules conformity*. Reusability of assets is the proportion of assets that are designed to be reusable. Coding rules conformity is the proportion of software modules that conform to given coding rules. In this context, the widespread use of automation development standards such as IEC 61131-3 and IEC 61499 means that reusability within software modules is *in-built*. Moreover, due to this high reusability of assets through the use of development standards, coding rule conformity is also built-in to a greater extent.

*3) Analysability:* Assessing a complete system for the impact from a localized change, identifying deficiencies, and finding parts that need to be modified are used in the analyzability context. *System log completeness* refers to the proportion of required logs that are actually recorded in a system. *Diagnosis function effectiveness* and *diagnosis function sufficiency* measure the proportion of implemented diagnostic functions that are useful, and the proportion of required diagnostics functions that are implemented, respectively. For this work's context, system log completeness is important not only for maintainability, but also for other qualities like security, reliability, and also for fault detection and reconfiguration. Diagnostics, while also important, have not found uniform use via development standards for industrial automation systems.

*4) Modifiability:* The ease at which a system or its parts can be modified without affecting overall product quality is closely linked with other sub-characteristics like modularity. *Modification efficiency* is measured as the average time taken per modification, normalized to the amount of modification time expected. *Modification correctness* measures the proportion of modifications that do not cause a quality degradation in the system, and *modification capability* is a time-boxed proportion of required modifications that could actually be implemented. In the paper's context, modification correctness is the most important measure, as it can be influenced through careful and systematic design. The other measures are historical, and can only be computed after the system has been operationalized.

*5) Testability:* The ease at which any part of a system can be tested to determine if it meets its target requirements and goals is also relevant. *Test function completeness* is a measure of how comprehensively tests are implemented within a system. *Autonomous testability* is the proportion of tests that can be run directly on individual components without depending on other (sub-) systems. It is linked closely with modularity measures. *Test restartability* is the proportion of tests that can be paused and restarted. In this work's context, *autonomous testability* is enabled through the use of development standards like IEC 61131-3 and IEC 61499. Software agent-control interfaces developed have well-defined test interfaces. *Test function completeness* is relevant but usually not a very achievable target given the highly compositional and distributed nature of such systems. However, aspects of the systems, such as human-machine interfaces, can be made more testable through provisions within accepted interfacing mechanisms like supervisory control. *Test restartability* is elusive in such dynamic systems, especially when agents are involved, as test playback to restart points is almost impossible.

**H. Portability**

Portability is used to judge on the level of transferability among different hardware, software or operational environments. To asses portability, ISO/IEC 25023 defines measures for *adaptability, installability* and *replaceability*. Although all of these metrics are relevant for industrial environments, the way some of them are measured needs to be in a more fine-grained context.

*1) Adaptability:* The degree of adaptability to different environments is assessed in different directions via *hardware environmental adaptability, system software environmental adaptability*, and *operational environment adaptability*. All three measures are considered as ratios of software functions that successfully operate on new conditions. Although for IA integration, all of the measures are relevant, one needs to explicitly define what the context and how much of the stack is attributed to the agent solution, e.g., if the OS on which running the agent platform is included, etc.

*2) Installability:* A successful installation is measured through time efficiency, and ease of installation customization. *Installation time efficiency* is highly relevant, however, this is measured against expected time, something that may be challenging to estimate. Also what the "installation" phase includes needs to be put into context, especially considering that there are common components (e.g., the agent platform, upon which variations of solutions may be installed); hence potentially this might be treated not at the system level, but down to individual components. Customization of the installation procedures is measured in *ease of installation* and is relevant, especially for agent solutions that feature self-X characteristics (configuration, adaption, etc.) and adjust their installation to their environments. This may also be linked with other phases like dealing with proper pre-operational correctness testing.

*3) Replaceability:* The level at which the agent-based solution can replace other solutions in the same environment is also relevant. All measurements are seen as suitable i.e. (i) *usage similarity*, which provides a metric of the functions that can be replaced (without workarounds or additional learning), (ii) *product quality equivalence*, which shows if the new product is better or equal than the previous, (iii) *functional inclusiveness* which reveals the percentage of functions that may give similar results with the old ones, and (iv) *data reusability/import capability* which assesses if the same data can be used as with the previous product. In industrial settings, if agent-based solutions are to replace existing monolithic products, the level of replaceability provides a highly relevant indicator, so that systems can be migrated to new ones, that perform similarly or better than the older counterparts.

## 3. Discussion

Integrating software agents and automation/control systems face broadly the same challenges as any other software projects, albeit there are context-specific distinctions which this paper attempts to highlight. ISO/IEC 25023 based measures for the eight characteristics (see section II) of quality software prove helpful towards comparing practices for integrating software agents and low-level automation functions.

Some of the measures, as shown in Table I, pose a very good fit overall, and as they are unambiguous and well-defined across any software system, such as the performance measures. However, some other measures have a low degree of suitability, since, although their corresponding characteristics can be relevant, the way ISO/IEC 25023 proposes to measure them may not be applicable or very difficult to realize in the IA integration context. For instance, some of these measures are vague or are designed to provide measurements for large software systems, capturing their system-level aspects. Hence, these need to be put into a more concrete perspective when discussing how to quantify them and clearly link them to an abstract model such as the one presented in Figure 1.

In some situations, while a measure proposed by the standard itself might be valid, the parameters used for that measurement may not fit the context. For instance, consider *cyclomatic complexity adequacy* which is a measure for modularity/maintainability. In a generic software system, the cyclomatic complexity of a module or sub-system must be kept reasonably low. However, in industrial automation systems in general, development standards are built strongly around the concept of high reuse through composition. In fact, the configuration of highly complex software made from reusable components is generally preferred over developing new, less complex blocks. This strategy keeps the number of reused software building blocks low, which then become easier to test comprehensively compared to having many blocks that are not reused as much. However, the cyclomatic complexity of blocks typically goes up, which goes against the measure as prescribed by the standard.

Some parameters defined in the ISO/IEC 25023 are quite similar and probably, selecting only a subset of them may be adequate to capture the necessary practice aspects for assessing it. As an example, the *failure rate* and the *MTBF* parameters are somehow similar in expressing the maturity of a software system and particularly the interface practice. Since they are correlated, it is usually preferable to use MTBF since the use of large numbers (e.g., 1000 hours) is more intuitive and easier to manage than very small numbers (e.g., 0.001 per hour). Another example is the *coupling* measure for modularity and the related measure *modification efficiency* for modifiability. Modification efficiency depends directly on coupling, that forms the basis for almost every other measure for maintainability, indicating that it is the most important measure to be captured for this characteristic.

The context in which the characteristics are assessed [10] also plays a key role. For instance, the meaning of failure needs to be agreed upon and understood in the specific context. In the scope of this work, a failure is related to a situation when the software agent cannot access to the low-level device to perform regular tasks due to a problem associated to the interface (including HLC or LLC API and communication infrastructure). Examples of failures can be broken connectivity, network problems, server non-responsiveness, etc.

Some measures are inherently difficult to quantify. Take as an example the "compatibility" context, where the integration practice needs to support excellently a single specific data exchange protocol and behave deterministically. However, given the large variety of protocols and data formats in shop-floor operations, such a practice will have low scores in compatibility (i.e, . on "data exchange protocol sufficiency") as this in the standard is simply defined as a ratio over the overall number of formats exchangeable with other systems.

Another pivotal issue is the coverage of the several life-cycle phases of the interface practices. Some measures are more focused in the design/testing phase, e.g., the *fault correction, coding rule conformity* and the *test coverage* measures. Since the main focus of the interface practices assessment is related to the operation phase, these measures are not that relevant to the interface practices context.

A question that arises is also which of these features are seen as "heavyweights" and therefore would have the most influence for a specific scenario. Surely there are no "one-size-fits-all" solutions, but nevertheless, when considering the specific characteristics of most industrial requirements some issues may emerge. For instance, in a small survey [13], carried out among the IEEE P2660.1 working group [18] experts, it emerged that *testability* is seen as much more important than others (e.g., the *user interface aesthetics*). However, as seen in Table I, the three proposed measures for assessing testability range from neutral, to good or very good. Hence some may be adopted as they are, while others (e.g., test restartability), might need to be tweaked to better match the agent context and the specific use-case requirements.

For different use cases or domains, the "heavyweightiness" of some characteristics impacts how it is perceived and how it can be captured in a representative way. For instance, a practice should be able to attain high efficiency when operating in isolation. However, such efficiency may not be representative of the normal operation of the system, as features, e.g., the agent's autonomous behavior, or run-time context (unforeseen side-effects among the services or infrastructure) may impose additional overhead and affect the metric. Therefore, adequate sampling of performance measurements with different software components and operational constellations [17] must be carefully carried out in order to produce representative results. Performance efficiency has been characterized before in respect to the agent platform in isolation [19], [20], but no efforts have been reported to systematically evaluate it at the interface between the agent platform and connected low-level automation functions. As such, a reassessment of the practice may be necessary, also when the practice is operational (in-use) in order to catch local setup effects that may impact some of the characteristics.

Apart from what is proposed by ISO/IEC 25023 and discussed so far, some additional measures can be used to more accurately capture some of the characteristics in this paper's context. For example, for Testability, we define the *reusability index* as the ratio of the number of blocks made from smaller reusable components to the total number of blocks in the system. A higher reusability index indicates that there is a smaller set of basic building blocks that can be tested in isolation. However, as systems and approaches become more intelligent and self-organized in order to meet stakeholder demands [4], measuring specific behaviors becomes challenging, as the system dynamically tries to adapt to meet the posed requirements.

The selection of measures may be differently perceived by the involved stakeholders, and therefore removing ambiguity about the context, the parameters, the way measurements are calculated, etc. could be beneficial. In addition, we need to point out that the practice assessment cannot be a one-time action. The measure assessment may change over successive system versions and therefore fluctuate over the lifetime of the practice. Hence, additional considerations to capture structural changes of the measures along evolution [21] may be needed.

## 4. Conclusions

ISO/IEC 25023 proposes several measures and ways to quantify them. When it comes to utilizing these measures in IA-based applications and more specifically to the integration with low-level automation functions, many but not all are meaningfully applicable. Some well-defined measures can of course be used and their quantification is appropriate, while some others might need to be tweaked to better reflect the context of IA. However, there are also other measures that have a poor relevance, and it does not make sense to utilize them. Independent of these results, there are also other issues that have emerged and may affect the utilization of these measures as discussed in section III, e.g., the lifecycle, the domain or use-case requirements, the exact context for multiagent systems and federated access to devices, etc. In addition, a challenging issue that needs to be addressed in the future is the automated testing and evaluation of the practices based on the criteria presented, as well as potential extensions of criteria themselves.